 \documentclass[manuscript]{emulateapj-rtx4}

\newcommand{\Ell}{E_\parallel}      
\newcommand{\rhoGJ}{\rho_{{\rm GJ}}}  
\newcommand{\rlc}{\varpi_{\rm LC}} 

\shorttitle{Hardening of Trailing Peak in Gamma-ray Pulsar Light Curve}
\shortauthors{Hirotani}

\begin{document}


\title{Pulsar Outer-gap Electrodynamics:
       Hardening of Spectral Shape in the Trailing Peak 
       in Gamma-ray Light Curve}


\author{Kouichi Hirotani\altaffilmark{1}}
\affil{Theoretical Institute for
       Advanced Research in Astrophysics (TIARA),
       Academia Sinica, Institute of Astronomy and Astrophysics (ASIAA),
       PO Box 23-141, Taipei, Taiwan}
\email{hirotani@tiara.sinica.edu.tw}

%

\altaffiltext{1}{Postal address: 
                 TIARA, Department of Physics, 
                 National Tsing Hua University,
                 101, Sec. 2, Kuang Fu Rd.,Hsinchu, Taiwan 300}




\begin{abstract}
The spectral characteristics of the pulsed gamma-ray emission
from outer-magnetospheric particle accelerators are investigated.
Either positrons or electrons are accelerated outwards by the
magnetic-field-aligned electric field to emit gamma-rays via
curvature process.
Since the particles move along relatively straight paths
in the trailing side of a rotating magnetosphere,
they attain higher Lorentz factors to emit more energetic gamma-rays
than those in the leading side.
It is first demonstrated that the cutoff energy of the curvature
radiation evolves with the rotation phase
owing to the variation of the curvature radii of the particle paths 
and maximizes at a slightly later phase of the trailing peak 
in the gamma-ray light curve.
\end{abstract}





\section{Introduction}
\label{sec:intro}
The {\it Fermi} Large Area Telescope (LAT) provides 
a wealth of new data on isolated, rotation-powered pulsars,
increasing the number of detected $\gamma$-ray pulsars
from seven to more than sixty 
(e.g., Abdo et al.~2010a; Abdo et al.~2010b; 
 Saz Parkinson et al. 2010).
The Astro-rivelatore Gamma a Immagini LEggero
({\it AGILE}) has also reported the detection of about twenty 
$\gamma$-ray pulsars 
(e.g., Pellizzoni et al.~2009; Pilia et al. 2010).
Since interpreting $\gamma$-rays is less ambiguous
compared with reprocessed, non-thermal X-rays,
the $\gamma$-ray pulsations observed from these objects
are particularly important as a direct signature of 
basic non-thermal processes in pulsar magnetospheres,
and should help to discriminate among different 
emission models.
In this letter, 
we thus focus on the spectral evolution of the
pulsed $\gamma$-rays and 
examine if the hardening of the trailing peak in the light curve,
a general property derived from these observations,
can be explained in the context of 
the outer-magnetospheric emission scenario.

The light curves and spectral evidence obtained by the observations
mentioned above,
suggest that the $\gamma$-ray pulsars have high-altitude emission
zones whose fan-like beams scan over a large fraction 
of the celestial sphere. 
Therefore, recent pulsar high-energy emission models adopt 
higher-altitude 
emission geometries.
There are three main scenarios in this approach:
the outer-gap (OG) model 
(Cheng et al.~1986a, b; Romani~1996; Cheng et al.~2000; 
 Hirotani~2006b, 2008; Takata et al.~2008; 
 Romani \& Watters~2010), 
the higher-altitude slot-gap model
(Muslimov \& Harding~2004),
and the pair-starved polar-cap model
(Frackowiak \& Rudak~2005; Harding et al.~2005; Venter et al.~2009).

In the present letter, we focus on the OG model,
in which pairs are created in the outer magnetosphere
mainly by photon-photon ($\gamma$-$\gamma$) pair production.
The produced pairs polarize owing to 
the magnetic-field-aligned electric field, $\Ell$.
If the rotation and magnetic axes reside in the same hemisphere,
a positive $\Ell$ is exerted to accelerate $e^+$'s (or $e^-$'s) 
outward (or inwards), increasing the real charge density outwards.
The inward-migrating, relativistic $e^-$'s radiate 
curvature $\gamma$-rays, 
some of which (nearly head-on) collide with the X-rays 
emitted from the NS surface to materialize as pairs.

In a pulsar magnetosphere, 
there is a surface called the \lq null surface', on which  
the Goldreich-Julian charge density
(Goldreich \& Julian~1969)
$\rhoGJ \equiv -\mbox{\boldmath$B$}\cdot \mbox{\boldmath$\Omega$}/(2\pi c)$ 
changes sign,
where $\mbox{\boldmath$B$}$ denotes the local magnetic field, 
$\mbox{\boldmath$\Omega$}$ the NS angular-velocity vector, and
$c$ the speed of light.
There is another characteristic surface called the \lq light cylinder' 
beyond which the co-rotational velocity exceeds $c$. 
The distance of the light cylinder from the rotation axis
is called the \lq light-cylinder radius',
$\rlc \equiv c/\Omega$, 
where $\Omega \equiv \vert \mbox{\boldmath$\Omega$} \vert$ denotes
the NS rotational angular frequency.
On each magnetic azimuthal angle $\varphi_\ast$ 
(measured around the magnetic axis counter-clockwise),
there is a magnetic field line that crosses the light cylinder
tangentially; 
they are called the \lq the last-open field lines'.
An OG is essentially located between the null surface and the 
light cylinder in the higher colatitudes 
(i.e., in the closer region to the magnetic axis) 
than the last-open field lines.

\section{Characteristic Energy of Curvature Radiation}
\label{sec:Ecutoff}
The characteristic energy of curvature emission
is given by (e.g., Rybicki \& Lightman 1979)
\begin{equation}
  E_{\rm c}
  = \frac32 \gamma^3 \frac{\hbar c}{\varrho_{\rm c}},
  \label{eq:Ec}
\end{equation}
where $\gamma$ refers to the Lorentz factor of
the electron or positron,
$\hbar$ the Planck constant divided by $2\pi$, and
$\varrho_{\rm c}$ the curvature radius of particle's
three-dimensional (3-D) motion. 
In the gap, the $e^\pm$'s achieve electrostatic force balance,
$ e\Ell= 2e^2\gamma^4 / (3\varrho_{\rm c}^2)$,
and saturate at the terminal Lorentz factor, 
\begin{equation}
  \gamma= \left( \frac{3}{2}
                 \frac{\varrho_{\rm c}{}^2}{e} \Ell
          \right)^{1/4},
  \label{eq:term}
\end{equation}
where $e$ designates the charge on the positron, and
$\Ell$ the electric field component projected along the
magnetic field line.
Combining equations~(\ref{eq:Ec}) and (\ref{eq:term}),
we obtain
\begin{equation}
  E_{\rm c}
  = \left(\frac32\right)^{7/4}
    \hbar c \varrho_{\rm c}{}^{1/2}
    \left(\frac{\Ell}{e}\right)^{3/4}.
  \label{eq:Ec2}
\end{equation}

Since $e^\pm$'s are saturated at $\gamma$,
the spectral density of their curvature emission
declines sharply as
$(E/E_{\rm c})^{1/2} \exp(-E/E_{\rm c})$
above the cutoff energy, $E \gg E_{\rm c}$.
Therefore, for the same $\Ell$,
equation~(\ref{eq:Ec2}) shows that the curvature spectrum has
a greater $E_{\rm c}$ for a greater $\varrho_{\rm c}$.

Since the pair production mainly takes place between
the inward $\gamma$-rays and the surface X-rays,
most of the pairs are produced in the
lower altitudes where the X-ray density is large and
the collisions take place mostly head-on.
Thus, outward-moving species (e.g., $e^+$'s)
run longer distances than inward ones within the gap,
leading to a stronger outward $\gamma$-ray flux than
the inward one.
Therefore, to elucidate the physical mechanism
that produces harder spectrum 
in the trailing peak of a $\gamma$-ray light curve,
we concentrate on the outward-moving particles
and the resultant outward emissions in \S~\ref{sec:Ecutoff}

\begin{figure}
 \includegraphics[angle=0,scale=0.50]{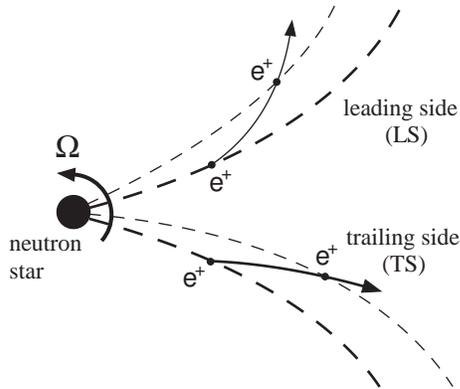}
\caption{
Schematic figure (top view) of particle motion in a rotating
pulsar magnetosphere.
The thick dashed lines represent the dipolar magnetic field lines
at some moment, and the thin dashed ones do the same field lines
at later rotation phase.
The thin and thick solid lines represent the paths of 
particles moving outwards in the leading side (LS) 
and trailing side (TS), respectively. 
\label{fig:Rc}
}
\end{figure}

To reveal the nature of curvature radiation,
we must consider the curvature radius of
actual particle motion in the 3-D pulsar magnetosphere.
Well inside the light cylinder,
the drift motion due to magnetic curvature, magnetic gradient, 
or time-varying electric field is small compared to the
$\mbox{\boldmath$E$}\times\mbox{\boldmath$B$}$ drift.
Thus, particles co-rotate with the magnetic field lines.
In the flaring field line geometry of a NS magnetosphere,
the path of outward-moving particles can be
illustrated as figure~\ref{fig:Rc}.
In the leading side, particle's path is bent towards the
rotational direction as the thin solid curve indicates.
However, in the trailing side, the path is straighter
as indicated by the thick solid curve.
Thus, outward-moving particles have greater curvature
radii in the trailing side than in the leading side.
This leads to a greater $E_{\rm c}$ in equation~(\ref{eq:Ec2}).
This is the qualitative reason why most of the
$\gamma$-ray pulsars detected with LAT exhibits harder
spectrum in the trailing light-curve peak than in the
leading peak.

Next, let us examine the distribution of the curvature radius
more quantitatively.
Due to strong relativistic beaming, 
curvature photons are emitted along the instantaneous particle 
velocity measured by the distant static observer (i.e., us).
In the polar coordinates, 
the orthonormal components of the particle velocity become
(Mestel et al. 1985; Camenzind 1986a,b)
\begin{equation}
  \frac{v^r}{c}
  = f_{\rm v} \frac{B^r}{B}, \quad
  \frac{v^{\hat\theta}}{c}
  = f_{\rm v} \frac{B^{\hat\theta}}{B}, \quad
  \frac{v^{\hat\phi}}{c}
  = f_{\rm v} \frac{B^{\hat\varphi}}{B} + \frac{\varpi}{\rlc},
  \label{eq:v3D}
\end{equation}
where $\gamma \gg 1$ is assumed and
\begin{equation}
  f_{\rm v} 
  \equiv -\frac{\varpi}{\rlc}\frac{B^{\hat\varphi}}{B}
         \pm \sqrt{1-\left(\frac{\varpi}{\rlc}\right)^2
                     \left(\frac{B_{\rm p}}{B}\right)^2};
  \label{eq:v3Df}
\end{equation}
$B_{\rm p}\equiv \sqrt{(B^r)^2+(B^{\hat\theta})^2}$,
$B$ denotes the strength of the magnetic field,
$\varpi$ the distance from the rotation axis.
Both ($v^r$,$v^{\hat\theta}$,$v^{\hat\phi}$) and
($B^r$,$B^{\hat\theta}$,$B^{\hat\varphi}$) are
measured by a distant static observer 
(i.e., in the inertial observers frame).
The upper sign of $f_{\rm v}$ (eq.~[\ref{eq:v3Df}]) 
corresponds to the outward motion along the magnetic field line,
whereas the lower sign to the inward motion.
These expressions are quite general and valid irrespective 
of the force balance on the particles as long as $\gamma \gg 1$,
because they are derived only from the Maxwell equations and the
frozen-in condition,
the latter of which is justified since we consider $\Ell \ll B$ 
(or equivalently, consider that the potential drop in the gap is 
 much small compared to the electro-motive force exerted on the 
 spinning NS surface).
For example, they are valid not only for $e^\pm$'s being 
accelerated without dissipation but also for those in the force 
balance between radiation-reaction and electro-static acceleration.
Note that equations~(\ref{eq:v3D}) and (\ref{eq:v3Df})
correspond to a generalization of \S~4.1 of Dyks et al. (2010)
into the higher altitudes.
Note also that equations~(\ref{eq:v3D}) and (\ref{eq:v3Df}) 
are applicable not only
within the light cylinder but also outside of it
and that $B^{\hat\varphi}<0$ holds in ordinary situations.

\begin{figure}
 \includegraphics[angle=0,scale=1.6]{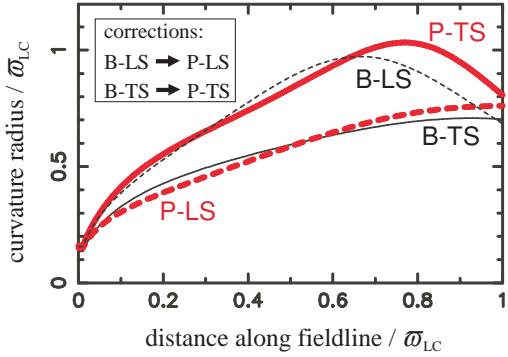}
\caption{
Distribution of the curvature radius, $\varrho_{\rm c}$ 
as a function of the distance along the magnetic field line.
Both the abscissa and the ordinate are normalized
by the light cylinder radius, $\rlc$.
The (red) thick solid curve (labeled as P-TS) denotes 
$\varrho_{\rm c}/\rlc$ of the actual particle path
in the TS (specifically, at $\varphi_\ast=-60^\circ$).
The (red) thick dashed curve (P-LS) denotes 
$\varrho_{\rm c}/\rlc$ of the actual particle path
in the LS ($\varphi_\ast=60^\circ$).
The (black) thin solid curve (B-TS) does $\varrho_{\rm c}/\rlc$ 
of the last-open magnetic field line in the TS
($\varphi_\ast=-60^\circ$), while
the (black) thin dashed curve (B-LS) that of the 
last-open magnetic field line in the LS
($\varphi_\ast=60^\circ$).
\label{fig:Rc2}
}
\end{figure}

Integrating equation~(\ref{eq:v3D}) along particle motion,
we obtain the path of the particle,
which allows us to compute $\varrho_{\rm c}$ as a function
of the distance $s$ along the field line.
Since the incorporation of a field-line deformation due to
magnetospheric currents is out of the scope of this letter,
we adopt a retarded, vacuum dipole field 
(Cheng et al. 2000) as an example.
In figure~\ref{fig:Rc2},
we present $\varrho_{\rm c}/\rlc$ as a function of $s$ for
a representative magnetic field line 
in the leading side (LS; $\varphi_\ast=+60^\circ$)
and that in the trailing side (TS; $\varphi_\ast=-60^\circ$),
where $\varphi_\ast$ refers to the magnetic azimuthal angle
measured around the magnetic axis counter-clockwise.
The magnetic inclination angle, $\alpha$, is assumed to be
$60^\circ$. 

In figure~\ref{fig:Rc2}, the (red) thick solid and dashed curves denote
$\varrho_{\rm c}/\rlc$ of the actual particle path,
whereas the (black) thin solid and dashed curves do 
$\varrho_{\rm c}/\rlc$ of the last-open magnetic field lines. 
The LS quantities are plotted with dashed curves,
while the TS quantities with solid curves.
It follows that $\varrho_{\rm c}/\rlc$ of the particle path
(labeled with P-TS) becomes greater than that of the field line
(B-TS) in the TS.
In the LS, on the contrary,
$\varrho_{\rm c}/\rlc$ of the particle path
(P-LS) becomes less than that of the field line (B-LS).
Note that B-TS is corrected to P-TS and
B-LS to P-LS, and that
the magnetic field lines (see B-LS vs. B-TS)
have greater $\varrho_{\rm c}$ in the LS whereas 
the particle paths (P-LS vs. P-TS)
have greater $\varrho_{\rm c}$ in the TS.
In previous pulsar emission models,
it has been phenomenologically assumed that the
particles emit curvature radiation along local
magnetic field lines and adopted B-LS and B-TS
as $\varrho_{\rm c}/\rlc$; 
thus, it has been impossible to
reproduce a harder spectrum in the TS than the LS.
Precisely speaking, 
they did considered the aberration of photon momentum 
to compute the emission direction from each point
(and to reproduce the observed pulse profiles),
but have computed $\varrho_{\rm c}$ from the magnetic field
geometry (as represented by the thin lines labeled with B-LS and B-TS).
Instead, we should take account of particle's' motion
with equation~(\ref{eq:v3D}) and adopt the correct $\varrho_{\rm c}$
(as represented by P-LS and P-TS);
then, we obtain a harder spectrum in the TS as observed
from most of the $\gamma$-ray pulsars.

\section{Self-consistent accelerator solution}
\label{sec:OG}
To precisely compute the phase-resolved spectrum,
we must solve $\Ell$ and the distribution functions
of $e^\pm$'s at each point in the 3-D pulsar magnetosphere,
together with the radiative transfer.
In this section, 
we thus solve the set of Maxwell and Boltzmann equations 
in the 3-D pulsar magnetosphere
in the open zone from the NS surface to the light cylinder
under appropriate boundary conditions
(see Hirotani~2006a, b for details).
To demonstrate general spectral properties,
we consider typical young NS parameters,
$P=0.1$~s, $\dot{P}=10^{-13}\,\mbox{s s}^{-1}$,
$kT_{\rm s}= 50$~eV, 
where $kT$ refers to the blackbody temperature of the whole NS 
thermal emission.
For this \lq example pulsar', we adopt $\alpha=60^\circ$.

\begin{figure}
 \includegraphics[angle=0,scale=2.0]{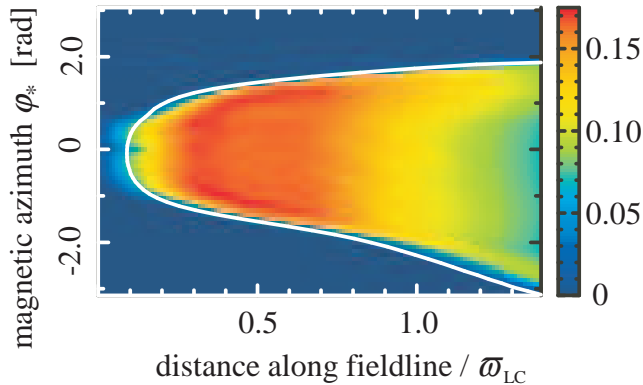} 
\caption{
Distribution of the gap meridional thickness projected 
on the last-open field line surface.
The white curve denotes the cross section between the
last-open field lines and the null surface.
The ordinate denotes the angle, $\varphi_\ast$ [rad], 
measured around the magnetic axis counter-clockwise.
The region $\varphi_\ast>0$ (or $\varphi_\ast<0$) is
referred to as the leading (or trailing) side.
\label{fig:hm}
}
\end{figure}

We present the distribution of the gap
in the magnetosphere by showing 
the gap fractional thickness, $h_{\rm m}$, projected on the
last-open field line surface in figure~\ref{fig:hm}.
If $h_{\rm m}=1.0$, it means that all the open fluxes threads
the OG.
On the other hand, if $h_{\rm m} \ll 1$,
only the fluxes very close to the last-open field lines
are active.
It follows that the gap exist mainly outside the null 
surface (white curve); thus,
the present solution coincidentally corresponds to a quantitative
extension of previous, phenomenological OG models.

\begin{figure}
 \includegraphics[angle=0,scale=2.00]{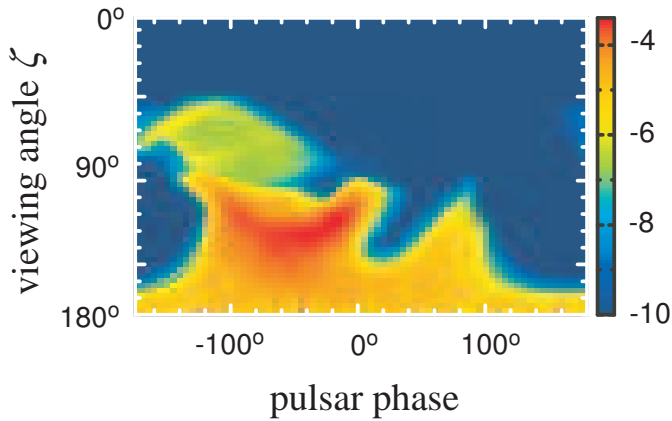} 
\caption{
Angular distribution of photon emission 
above 89~MeV 
($\log_{10}$ of flux per phase in 
 $\mbox{MeV\,s}^{-1}\,\mbox{cm}^{-2}\,\mbox{deg}^{-1}$ unit)
from the outer gap connected to the north magnetic pole.
The phase $0^\circ$ (in the abscissa)
corresponds to the neutron-star rotation phase
at which the photon emitted from the north polar cap
arrives the observer.
The ordinate denotes the observer's viewing angle $\zeta$ 
with respect to the rotation axis.
\label{fig:map}
}
\end{figure}

We next present the distribution of photon emission
as a function of NS rotational phase and the observers'
viewing angle $\zeta$ in figure~\ref{fig:map},
where the distance is assumed to be 1~kpc.
Since only the photons emitted from the OG connected to the
north magnetic pole are plotted, 
photons propagating into the southern hemisphere ($\zeta>90^\circ$)
are mostly emitted by outward-moving charges, whereas
photons into $\zeta<90^\circ$ are by inward ones.
Thus, we find that the outward photon flux dominates the inward one.

\begin{figure}
\epsscale{1.0}
  \includegraphics[angle=0,scale=1.20]{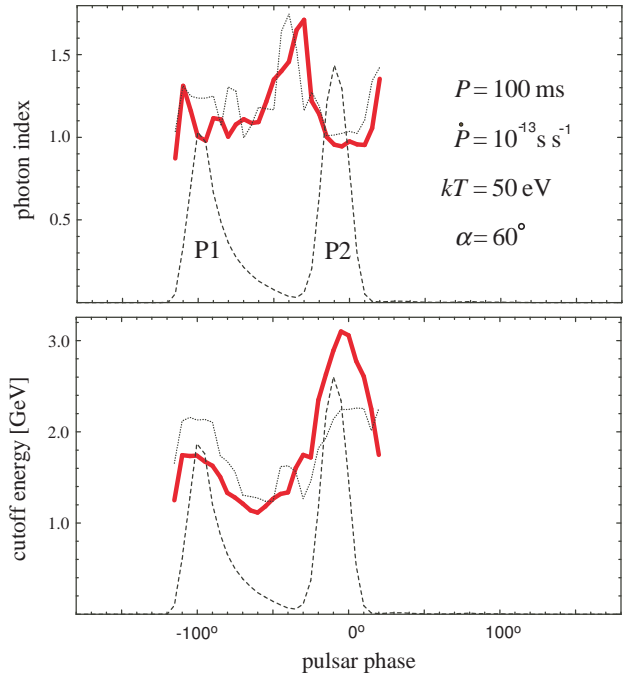}
 \caption{
 Evolution of the photon index $\Gamma$ (top panel) and
 the cutoff energy $E_{\rm c}$ (bottom panel) with the rotational phase,
 is depicted by the (red) thick solid curve,
 for an example pulsar with $P=0.1$~s, 
 $\dot{P}=10^{-13}\,\mbox{s s}^{-1}$,
 $kT=50$~eV, and $\alpha=60^\circ$.
 Observer's viewing angle is assumed to be $\zeta=115^\circ$.
 For comparison, $\Gamma$ and $E_{\rm c}$
 that would be obtained if we adopted the curvature radii
 of the magnetic field lines (instead of the particle paths)
 are depicted by the thin dotted curves.
 Pulse profile is overlaid as the dashed curves (in arbitrary unit).
 \label{fig:spc}
 }
 \end{figure}

We can find the pulse profiles by slicing
across the photon intensity plot in figure~\ref{fig:map}
at a constant $\zeta$.
As an example, we choose $\zeta=115^\circ$ and show
the pulse profile in figure~\ref{fig:spc} 
as the dashed line (in arbitrary unit);
P1 corresponds to the leading peak, while P2 the trailing peak.
One full NS rotation period is depicted in the abscissa.

Finally and most importantly,
we examine the evolution of spectral properties with phase.
To this aim, we fit the solved differential photon flux
($\mbox{photons cm}^{-2}\,\mbox{s}^{-1}\,\mbox{MeV}^{-1}$)
with a power law with exponential cutoff shape given by
\begin{equation}
  \frac{dN}{dE}= K E^{-\Gamma}
                   \left( -\frac{E}{E_{\rm c}} \right)
\end{equation}
in each phase bin with $1/72$ (i.e., $5^\circ$) interval,
where both $E$ and $E_{\rm c}$ are in GeV unit and 
refer to the photon energy and the cutoff energy, respectively;
$\Gamma$ is the photon index.
The best fit $\Gamma$ and $E_{\rm c}$ are depicted in
figure~\ref{fig:spc}.
The (red) thick solid curve in the top panel denotes
the evolution of $\Gamma$ with phase, while 
that in the bottom panel that of $E_{\rm c}$.
For comparison, $\Gamma$ and $E_{\rm c}$ that would be
obtained if we adopted $\rho_{\rm c}$ of the magnetic field lines
(instead of that of the particle path)
are also depicted by thin dotted curves.
It follows that the photon index remains roughly constant
during the bright phase,
while the cutoff energy sharply peaks in the trailing peak
as expected from the analytic argument in \S~\ref{sec:Ecutoff}.
Note that only comparable $E_{\rm c}$ would be obtained 
in the two light-curve peaks 
if we adopted $\rho_{\rm c}$ of the magnetic field lines
(as in previous pulsar emission models).
The conclusion of greater $E_{\rm c}$ in the trailing peak
is, in deed, consistent with
the spectral properties of most of the
$\gamma$-ray pulsars detected with LAT 
(for PSR~J2021+3651, see Abdo et al. 2009;
 for the Crab pulsar, see Abdo et al. 2010b;
 for the Vela pulsar, see Abdo et al. 2010c;
 for PSR~J1057--5226, PSR~J1709--4429, PSR~J1952+3252,
 see Abdo et al. 2010d;
 for the Geminga pulsar, see Abdo et al. 2010e).

\section{Discussion}
\label{sec:discussion}
By examining the actual paths of outward-moving charged
particles in a 3-D pulsar magnetosphere, 
we demonstrated that the curvature radii of their paths become larger
in the trailing side (TS) than in the leading side (LS)
in \S~\ref{sec:Ecutoff}.
Applying this result to the curvature radiation formula
(\S~\ref{sec:intro}),
we first gave a physical explanation why
the trailing peak generally exhibits a harder spectrum 
than the leading peak in $\gamma$-ray light curves.
To confirm this analytical prediction,
we solved the set of Maxwell and Boltzmann equations
in the 3-D pulsar magnetosphere in \S~\ref{sec:OG}, 
finding that the obtained emissivity distribution
does result in a greater cutoff energy in the trailing peak
than in the leading peak.
Although the discussion in ~\S~\ref{sec:OG} is limited to 
the OG model, 
the same argument can be, in fact, readily applied to 
other higher-altitude emission models,
as long as the emission is due to the curvature process.
What we have to change is to amend $\Ell$ 
(instead of what is obtained in the OG model) 
as a function of position in the 3-D pulsar magnetosphere.
Unless $\Ell$ becomes substantially greater in the LS than in the TS,
the same conclusion will be derived,
which is, in deed, expected both in the slot-gap and in the
pair-starved polar-cap models.

In figure~\ref{fig:spc},
$\Gamma$ peaks at pulsar phase $\sim -30^\circ$.
However, this peaking structure of $\Gamma$ appears at 
small-flux phase, and hence is vulnerable to the choice of $\zeta$.
For example, it disappears for $\zeta>120^\circ$.
On the contrary, the peaking structure of $E_{\rm c}$ 
at a slightly later phase of the trailing peak, 
is generally obtained in a wide parameter range 
of ($P$,$\dot{P}$,$kT$,$\alpha$) and $\zeta$.
Although not all the parameter space has not been investigated,
the conclusion that the trailing peak has a greater cutoff energy
than in other phases, appears to be universal. 

In this letter, we have adopted a retarded, vacuum dipole magnetic field. 
However, in general, we should take account of the deformation of 
the field lines due to the magnetospheric current
(Contopoulos et al. 1999; Gruzinov 2005; 
 Spitkovsky 2006; Bai \& Spitkovsky 2010a, b; 
 Contopoulos \& Kalapotharakos 2010).
The effect of deformation works to straighten the field lines
in the outer magnetospheres, pushing the OG position
towards the light cylinder particularly in the TS; 
for example, if $\alpha=60^\circ$,
the OG within the region $\varphi_\ast < -60^\circ$
is suggested to be less active due to the position shift towards the
light cylinder.
Nevertheless, the self-consistent OG solution presented 
in \S~\ref{sec:OG} shows 
that the pair production is not, any way, self-sustained in
$\varphi_\ast < -60^\circ$,
resulting in a negligible emission from this region.
For example, the trailing peak (around $-10^\circ$ in pulsar phase)
is mostly formed by the emission from 
$-60^\circ < \varphi_\ast < -30^\circ$.
In addition, the specific intensity towards $\zeta \sim 115^\circ$ 
direction will not be strongly affected by the straightening, 
which follows from panel (d) of figure~8 in Bai and Spitkovsky (2010b).
On these grounds, we consider that the conclusions in the present letter 
are relatively less vulnerable to the correction of the field-line 
geometry due to magnetospheric currents.


We can further apply the present numerical method to any
rotation-powered pulsars. 
In subsequent papers, we will demonstrate that the numerical
solutions are consistent with various observations
of individual pulsars,
such as the energy fluxes, pulse profiles, and
phase-resolved spectra. 
To this aim, it is essential to 
self-consistently solve the screening processes 
due to the polarization of the produced pairs,
together with the distribution functions of $e^\pm$'s and 
photons.
This can be achieved only by solving the Maxwell and
Boltzmann equations simultaneously in the 3-D pulsar magnetosphere.

\acknowledgments
 This work is supported by the Theoretical Institute for
 Advanced Research in Astrophysics (TIARA) operated under 
 Academia Sinica,
the National Science Council Excellence Projects program 
in Taiwan administered through grant number 
NSC 96-2752-M-007-006-PAE,
 and the Formosa Program between National Science Council  
 in Taiwan and Consejo Superior de Investigaciones Cientificas
 in Spain administered through grant number 
 NSC100-2923-M-007-001-MY3.

\acknowledgments

\end{document}